\documentclass[aps,showpacs,msmath,amssymb,prl,twocolumn]{revtex4}
\usepackage{epsfig}

\newcommand\bb[1] {\mbox{\boldmath{$#1$}}}
\newcommand\del{\bb{\nabla}} 
\newcommand\bcdot{\bb{\cdot}}
\newcommand\btimes{\bb{\times}} 
\newcommand\real{{\rm Re}} 

\begin{document}

\title{\textsc{A Local Model for Angular Momentum Transport in
Accretion Disks \\ Driven by the Magnetorotational Instability}}
\author{Martin E. Pessah$^{1,2}$\email{mpessah@as.arizona.edu},
Chi-kwan Chan$^2$, and Dimitrios Psaltis$^{2,1}$} \affiliation{
$^{1}$Department of Astronomy, University of Arizona, Tucson, AZ 85721
\\ $^{2}$Department of Physics, University of Arizona, Tucson, AZ 85721}

\email{mpessah@as.arizona.edu}

\begin{abstract}
  We develop a local model for the exponential growth and saturation
  of the Reynolds and Maxwell stresses in turbulent flows driven by
  the magnetorotational instability. We first derive equations that
  describe the effects of the instability on the growth and pumping of
  the stresses. We highlight the relevance of a new type of
  correlations that couples the dynamical evolution of the Reynolds
  and Maxwell stresses and plays a key role in developing and
  sustaining the magnetorotational turbulence. We then supplement
  these equations with a phenomenological description of the triple
  correlations that lead to a saturated turbulent state. We show that
  the steady-state limit of the model describes successfully the
  correlations among stresses found in numerical simulations of
  shearing boxes.
\end{abstract}

\maketitle


Since the early days of accretion disk theory, it has been recognized
that molecular viscosity cannot account for a number of observational
properties of accreting objects.  Shakura and Sunyaev \cite{SS73}
introduced a parametrization of the shear stress that has been widely
used since.  Much of the success of their model lies on that many disk
observables are determined mostly by energy balance and depend weakly
on the adopted prescription \cite{BP99}.  However, this
parametrization leaves unanswered fundamental questions on the origin
of the anomalous transport and its detailed characteristics.

Strong support for the relevance of magnetic fields in accretion disks
arose with the realization that differentially rotating flows with
radially decreasing angular velocities are unstable when threaded by
weak magnetic fields \cite{BH91-98}.  Since the discovery of this
magnetorotational instability (MRI), a variety of local
\cite{HGB95-96,Sanoetal04} and global \cite{H00-01,SP01,HK01}
numerical simulations have confirmed that its long-term evolution
gives rise to a sustained turbulent state and outward angular momentum
transport. However, global simulations also demonstrate that angular
momentum transfer in turbulent accretion disks cannot be
adequately described by the Shakura \& Sunyaev prescription.  In
particular, there is evidence that the turbulent stresses are not
proportional to the local shear \cite{ABL96} and are not even
determined locally \cite{Armitage98}.

Some attempts to eliminate these shortcomings have been made within
the formalism of mean-field magnetohydrodynamics \cite{KR80}.
This has been a fruitful approach for modeling the growth of mean
magnetic fields in differentially rotating media \cite{BS05}. However,
it has proven difficult to use a dynamo model to describe the
transport of angular momentum in these systems \cite{Blackman01,
Brandenburg05}. This is especially true in turbulent flows driven by
the MRI, where the fluctuations in the magnetic energy are larger than
the mean magnetic energy, and the turbulent velocity and magnetic
fields evolve simultaneously \cite{BH91-98}.

In order to overcome this difficulty various approaches have been
taken in different physical setups.  Blackman and Field~\cite{BF02}
derived a dynamical model for the nonlinear saturation of helical
turbulence based on a damping closure for the electromotive force in
the absence of shear.  On the other hand, Kato and Yoshizawa
\cite{KY93-95,Kato98} and Ogilvie~\cite{Ogilvie03} derived a set of
closed dynamical equations that describe the growth and saturation of
the Reynolds and Maxwell stresses in shearing flows in the absence of
mean magnetic fields.

In this {\em Letter}, we relax some of the key assumptions made in
previous works and develop the first local model for the dynamical
evolution of the Reynolds and Maxwell tensors in turbulent magnetized
accretion disks that incorporates explicitly the MRI as their source.

The equation describing the dynamical evolution of the mean angular
momentum density, $\bar{l}$, of a fluid element in an accretion disk
with tangled magnetic fields is
\begin{equation}
\label{eq:angular_momentum_mean_weak}
\partial_t \bar l + \del \bcdot (\bar l \bar{\bb{v}}) =  - \del \bcdot
(r \bar{\bb{\mathcal F}}) \,.
\end{equation}
Here, the over-bars denote properly averaged values, $\bar{\bb{v}}$ is
the mean flow velocity, and the vector $\bar{\bb{\mathcal F}}$
characterizes the flux of angular momentum. Its radial component,
$\bar{\mathcal F}_r \equiv \bar{R}_{r\phi} - \bar{M}_{r\phi}$, is
equal to the total stress acting on a fluid element.  The stresses due
to correlations in the velocity field, $\bar{R}_{r\phi} \equiv \langle
\rho \delta\!v_r \delta\!v_\phi\rangle$, and magnetic field
fluctuations, $\bar{M}_{r\phi} \equiv \langle\delta\!B_r
\delta\!B_\phi\rangle/4\pi$, are the Reynolds and Maxwell stresses,
respectively. A self-consistent accretion disk model based on the
solution of the equations for the mean velocities requires a closed
system of equations for the temporal evolution of these mean stresses.

In a recent paper \cite{PCP06}, we identified the signature of the
magnetorotational instability in the mean Maxwell and Reynolds
stresses due to correlated fluctuations of the form $\delta
\bb{v}=[\delta v_r(t,z), \delta v_\phi(t,z), 0]$ and $\delta
\bb{B}=[\delta B_r(t,z), \delta B_\phi(t,z), 0]$ in an incompressible,
cylindrical, differentially rotating flow, threaded by a mean vertical
magnetic field, $\bar{B}_z$.  We demonstrated that a number of
properties of the mean stresses during the initial phase of
exponential growth of the MRI are approximately preserved in the
saturated state reached in local three-dimensional numerical
simulations.  As a first step in our calculation, we aim to derive
dynamical equations for the mean stresses that describe this result.

In order to work with dimensionless variables we consider the
characteristic time- and length-scales set by $1/\Omega_0$ and
$\bar{v}_{{\rm A} z}/\Omega_0$. Here, $\Omega_0$ and $\bar{v}_{{\rm A}
z}=\bar{B}_z/\sqrt{4\pi\rho_0}$ denote the local values of the angular
frequency and the (vertical) Alfv\'en speed in the disk with local
density $\rho_0$ at the fiducial radius $r_0$.  The equations
governing the local dynamics of the (dimensionless) MRI-driven
fluctuations in Fourier space, are then given by \cite{PCP06}
\begin{eqnarray}
\label{eq:ft_vx_nodim}
\partial_t \hat{\delta v_r} &=& 2 \hat{\delta v_\phi} + 
 ik_n \hat{\delta b_r} \,,  \\
\label{eq:ft_vy_nodim}
\partial_t \hat{\delta v_\phi} &=& (q-2) \hat{\delta v_r} + 
 ik_n \hat{\delta b_\phi} \,, \\
\label{eq:ft_bx_nodim}
\partial_t \hat{\delta b_r} &=&  ik_n \hat{\delta v_r} \,,  \\
\label{eq:ft_by_nodim}
\partial_t \hat{\delta b_\phi} &=& - q \hat{\delta b_r} + 
 i k_n \hat{\delta v_\phi} \,,
\end{eqnarray}
where $\hat{\delta v_i}$ and $\hat{\delta b_i}$ stand for the Fourier
transform of the dimensionless physical fluctuations $\delta v_i
/\bar{v}_{{\rm A} z}$ and $\delta B_i/\bar{B}_z$.  The wavenumber,
$k_n$, denotes the mode with $n$ nodes in the vertical direction and
the parameter $q\equiv-d\ln\Omega/d\ln r|_{r_0}$ is a measure of the
local shear.

Using the fact that the modes with vertical wavevectors dominate the
fast growth driven by the MRI, we obtain a set of equations to
describe the initial exponential growth of the Reynolds and Maxwell
stresses. We start from the equations for the fluctuations
(\ref{eq:ft_vx_nodim})--(\ref{eq:ft_by_nodim}) and use the fact that
the mean value of the product of two functions $f$ and $g$, with zero
means, is given by \cite{PCP06}
\begin{eqnarray}
\label{eq:mean_ft}
\langle f g\rangle (t) \equiv  2 \sum_{n=1}^{\infty} 
\; {\rm Re}[\,\hat{f}(k_n,t)\, \hat{g}^*(k_n,t)\,] \,.
\end{eqnarray}
By combining different moments of equations
(\ref{eq:ft_vx_nodim})--(\ref{eq:ft_by_nodim}) we obtain the
dimensionless set
\begin{eqnarray}
\label{eq:mean_Rrr}
\partial_t \bar{R}_{rr} &=& 4 \bar{R}_{r\phi}+2 \bar{W}_{r\phi}
\,, \\
\label{eq:mean_Rrphi}
\partial_t \bar{R}_{r\phi} &=& 
(q-2) \bar{R}_{rr} + 2 \bar{R}_{\phi\phi} - \bar{W}_{rr} +
\bar{W}_{\phi\phi} \,, \\
\label{eq:mean_Rphiphi}
\partial_t \bar{R}_{\phi\phi} &=& 2(q-2) \bar{R}_{r\phi}
- 2 \bar{W}_{\phi r} \,, \\ 
\label{eq:mean_Mrr}
\partial_t \bar{M}_{rr} &=& -2 \bar{W}_{r\phi}
\,, \\
\label{eq:mean_Mrphi}
\partial_t \bar{M}_{r\phi} &=& -q \bar{M}_{rr}
+ \bar{W}_{rr} - \bar{W}_{\phi\phi}
\,, \\
\label{eq:mean_Mphiphi}
\partial_t \bar{M}_{\phi\phi} &=& -2q \bar{M}_{r\phi}
+2 \bar{W}_{\phi r}
\,,
\end{eqnarray}
where we have defined the tensor $\bar{W}_{ij}$ with components
\begin{eqnarray}
\bar{W}_{rr} &\equiv& \langle\delta v_r \delta j_r\rangle =
\,\,\,\,\, \langle\delta b_\phi \delta \omega_\phi\rangle\,, 
\label{eq:W_rr}\\
\bar{W}_{r\phi} &\equiv& \langle \delta v_r \delta j_\phi \rangle =
-\langle\delta b_r \delta \omega_\phi\rangle\,, \\
\bar{W}_{\phi r} &\equiv& \langle \delta v_\phi \delta j_r \rangle =
-\langle\delta b_\phi \delta \omega_r\rangle\,, \\
\bar{W}_{\phi \phi} &\equiv& \langle \delta v_\phi \delta j_{\phi}\rangle =
\,\,\,\,\, \langle\delta b_r \delta \omega_r\rangle\,.
\label{eq:W_pp}
\end{eqnarray}
Here, $\delta j_i$ and $\delta \omega_i$, for $i=r,\phi$, stand for
the components of the induced current $\delta \bb{j} = \del \btimes
\delta \bb{b}$ and vorticity $\delta \bb{\omega} = \del \btimes \delta
\bb{v}$ fluctuations. Note that the components of the tensor
$\bar{W}_{ij}$ are defined in terms of the correlations between the
velocity and current fields. However, for the case under
consideration, these are identical to the corresponding correlations
between the magnetic field and vorticity fluctuations
(eqs.~[\ref{eq:W_rr}]-[\ref{eq:W_pp}]).

Equations~(\ref{eq:mean_Rrr})--(\ref{eq:mean_Mphiphi}) show that the
MRI-driven growth of the Reynolds and Maxwell tensors can be described
formally {\it only} via the correlations $\bar{W}_{ij}$ that connect
the equations for their temporal evolution.  Note that, in contrast to
the correlations $\langle \delta v_i \delta \omega_k\rangle$ and
$\langle \delta b_i \delta j_k \rangle$, that appear naturally in
helical dynamo modeling and transform as tensor densities (i.e., as
the product of a vector and an axial vector), the correlations
$\langle \delta v_i \delta j_k\rangle$ and $\langle \delta b_i \delta
\omega_k \rangle$ transform as tensors.  Moreover, the tensor
$\bar{W}_{ij}$ cannot be recast in terms of the cross-helicity tensor
$\bar{H}_{ij}\equiv\langle\delta v_i \delta b_j\rangle$ because, for
the unstable MRI modes, the ratio $\bar{H}_{ij}/\bar{W}_{ij}$
approaches zero at late times.  Neither can $\bar{W}_{ij}$ be
expressed in terms of the turbulent electromotive force $\langle\delta
\bb{v} \btimes \delta \bb{b}\rangle$, as the latter vanishes under our
set of assumptions, implying that no mean magnetic field is generated.

In order to describe the MRI-driven exponential growth of the stresses
in equations~(\ref{eq:mean_Rrr})--(\ref{eq:mean_Mphiphi}) we need to
write an additional set of dynamical equations for the evolution of
the tensor $\bar{W}_{ij}$. Using appropriate combinations of different
moments of equations (\ref{eq:ft_vx_nodim})--(\ref{eq:ft_by_nodim}) we
obtain
\begin{eqnarray}
\label{eq:mean_Wrr}
\partial_t \bar{W}_{rr} \!\!&=&\!\! q \bar{W}_{r \phi} + 2 \bar{W}_{\phi r} + 
(\bar{k}^R_{r\phi})^2 \bar{R}_{r\phi} - (\bar{k}^M_{r\phi})^2 \bar{M}_{r\phi}\,, \\
\label{eq:mean_Wrphi}
\partial_t \bar{W}_{r\phi} \!\!&=&\!\! 2 \bar{W}_{\phi \phi} -  
(\bar{k}^R_{rr})^2 \bar{R}_{rr} + (\bar{k}^M_{rr})^2 \bar{M}_{rr}\,, \\
\label{eq:mean_Wphir}
\partial_t \bar{W}_{\phi r} \!\!&=&\!\! (q-2) \bar{W}_{rr} + q \bar{W}_{\phi\phi} +
(\bar{k}^R_{\phi\phi})^2 \bar{R}_{\phi\phi} - (\bar{k}^M_{\phi\phi})^2 \bar{M}_{\phi\phi}\,,\ \ \ \ \ \ \\
\label{eq:mean_Wphiphi}
\partial_t \bar{W}_{\phi\phi} \!\!&=&\!\! (q-2) \bar{W}_{r\phi} - 
(\bar{k}^R_{r\phi})^2 \bar{R}_{r\phi} + (\bar{k}^M_{r\phi})^2 \bar{M}_{r\phi}\,,
\end{eqnarray}
where we have defined the set of mean wavenumbers 
\begin{eqnarray}
(\bar{k}^R_{ij})^2 \equiv
\frac{\displaystyle \sum_{n=1}^{\infty} k_n^2 \real[\hat{\delta v_i}
  \hat{\delta v_j}\!\!^*]}
{\displaystyle \sum_{n=1}^{\infty} \real[\hat{\delta v_i}
  \hat{\delta v_j}\!\!^*]} \,,\,
(\bar{k}^M_{ij})^2 \equiv
\frac{\displaystyle \sum_{n=1}^{\infty} k_n^2 \real[\hat{\delta b_i} \hat{\delta b_j}\!\!^*]}
{\displaystyle \sum_{n=1}^{\infty} \real[\hat{\delta b_i}
  \hat{\delta b_j}\!\!^*]} \,. \ \ \ \  \nonumber
\label{eq:k_ave}
\end{eqnarray}

The system of equations~(\ref{eq:mean_Rrr})--(\ref{eq:mean_Mphiphi})
and (\ref{eq:mean_Wrr})--(\ref{eq:mean_Wphiphi}) describes the
temporal evolution of the stresses during the exponential growth of
the MRI in a way that is formally correct, with no approximations.
Motivated by the similarity between the ratios of the stresses during
the exponential growth of the MRI and during the saturated turbulent
state~\cite{PCP06}, we propose to use the right-hand sides of
equations~(\ref{eq:mean_Rrr})--(\ref{eq:mean_Mphiphi}) and
(\ref{eq:mean_Wrr})--(\ref{eq:mean_Wphiphi}) as a local model for the
source of turbulence in MRI-driven magnetohydrodynamic flows.  Of
course, in the turbulent regime, the various average wavenumbers
$\bar{k}^{R,M}_{ij}$ will depend on the spectrum of velocity and
magnetic field fluctuations. As the lowest order model for these
wavenumbers we choose
\begin{eqnarray}
\label{eq:kbar}
(\bar{k}^{R,M}_{ij})^2 = \zeta^2 k^2_{\rm max}
= \zeta^2 \left(q - \frac{q^2}{4}\right) \quad \textrm{for} \quad i,j =
r,\phi \,, \ \ \ \
\end{eqnarray}
where $\zeta$ is a parameter of order unity and $k_{\rm max}$
corresponds to the wavenumber at which the growth rate of the {\it
fluctuations} reaches its maximum value, $\gamma_{\rm max} \equiv
q/2$.

\begin{figure}[t]
\epsfig{file=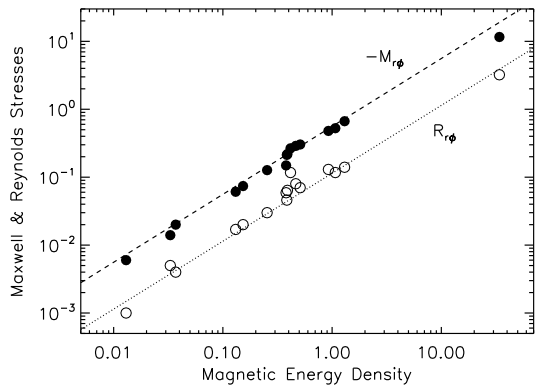, width=0.825\columnwidth,trim=0 10 0 10}
\caption{\footnotesize Correlations between the Maxwell and Reynolds
    stresses and mean magnetic energy density at saturation in
    MRI-driven turbulent shearing boxes \cite{HGB95-96}. The lines
    show the result obtained with our model in the asymptotic limit
    for $\zeta=0.3$.}
\label{fig:stresses_vs_emagnetic}
\end{figure}

By construction, the set of
equations~(\ref{eq:mean_Rrr})--(\ref{eq:mean_Mphiphi}) and
(\ref{eq:mean_Wrr})--(\ref{eq:mean_Wphiphi}) leads to the expected
exponential growth of the mean stresses driven by modes with
wavevectors perpendicular to the disk midplane. These modes, however,
are known to be subject to parasitic instabilities, which transfer
energy to modes in the perpendicular ($k_r,k_\phi$) directions
\cite{GX94}. The initial fast growth experienced by the stresses will
eventually be slowed down by the combined effects of non-linear
couplings between modes and of dissipation at the smallest scales of
interest.

The terms accounting for these interactions would appear in the
equations for the stresses as triple correlations between components
of the velocity and magnetic fields, i.e., they would be of the form
$\langle\delta v_i \delta v_j \delta b_k\rangle$ and $\langle\delta
v_i \delta b_j \delta b_k\rangle$.  These types of non-linear terms
have also been considered by Ogilvie \cite{Ogilvie03}, who proposed
scalings of the form
\begin{equation}
\langle\delta v_i \delta v_j \delta b_k\rangle
\sim \langle\delta v_i \delta v_j\rangle \, \langle\delta b_k \delta
b_k\rangle^{1/2} \sim \bar{M}_{kk}^{1/2}\bar{R}_{ij} \,.
\end{equation}

In the absence of a detailed model for these correlations and
motivated again by the similarity of the stress properties during the
exponential growth of the MRI and the saturated turbulent
state~\cite{PCP06}, we introduce a phenomenological description of the
non-linear effects on the evolution of the various stresses.  In
particular, denoting by $\bar{X}_{ij}$ the $ij$-component of any one
of the three tensors, i.e., $\bar{R}_{ij}$, $\bar{M}_{ij}$, or
$\bar{W}_{ij}$, we add to the equation for the temporal evolution of
that component the sink term 
\begin{equation}
\left.\frac{\partial\bar{X}_{ij}}{\partial t}\right\vert_{\rm sink} \equiv
-\sqrt{\frac{\bar{M}}{\bar{M}_0}} \bar{X}_{ij} \,.
\label{eq:sink}
\end{equation}
Here, $\bar{M}/2 = (\bar{M}_{rr} + \bar{M}_{\phi\phi})/2$ is the mean
magnetic energy density in the fluctuations and $\bar{M}_0$ is a
parameter.

Adding the sink terms to the system of
equations~(\ref{eq:mean_Rrr})--(\ref{eq:mean_Mphiphi}) and
(\ref{eq:mean_Wrr})--(\ref{eq:mean_Wphiphi}) leads to a saturation of
the stresses after a few characteristic timescales, {\em preserving\/}
the ratio of the various stresses to the value determined by the
exponential growth due to the MRI and characterized by the parameter
$\zeta$. These non-linear terms dictate only the saturated level of
the mean magnetic energy density according to $\lim_{t\rightarrow
\infty}{\bar{M}}= \Gamma^2 \, \bar{M_0}$, where $\Gamma$ is the growth
rate for the {\it stresses}, which, in the case of a Keplerian disk,
with $q=3/2$, is given by $\Gamma^2=2\left[\sqrt{1 + 15\zeta^2} -
\left(1+15\zeta^2/8\right) \right]$.

\begin{figure}[t]
\epsfig{file=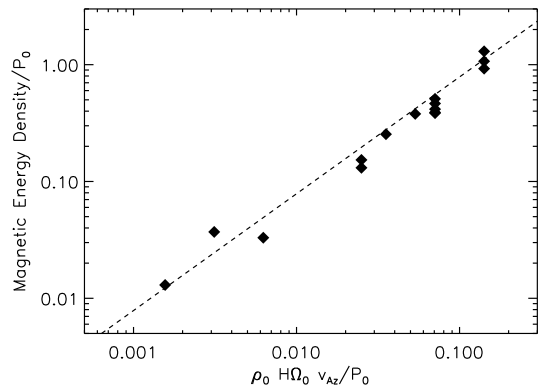, width=0.825\columnwidth,trim=0 10 0 10}
\caption{\footnotesize The mean magnetic energy density in terms of a
  saturation predictor found at late times in numerical simulations of
  MRI-driven turbulence \cite{HGB95-96}. The line shows the result
  obtained with our model in the asymptotic limit for
  $\zeta=0.3$~and~$\xi=11.3$.}
\label{fig:saturator_predictor}
\end{figure}

We infer the dependence of the energy density scale $\bar{M}_0/2$ on
the four characteristic scales in the problem $\Omega_0$, $H$ (the
vertical length of the box), $\rho_0$, and $\bar{v}_{{\rm A} z}$ using
dimensional analysis. We obtain $\bar{M}_0/2 \propto \rho_0 H^\delta
\Omega_0^{\delta} \bar{v}_{{\rm A}z}^{2-\delta}$ which leads, with the
natural choice $\delta=1$, to $\bar{M}_0/2 \equiv \xi \rho_0 H\Omega_0
\bar{v}_{{\rm A} z}$, where we show the dimensional quantities
explicitly and introduce the parameter $\xi$.

Our expression for $\bar{M}_0$ describes the same scaling between the
magnetic energy density during saturation and the various parameters
characterizing the disk found in a series of shearing box simulations
threaded by a finite vertical magnetic field and with a Keplerian
shearing profile \cite{HGB95-96}.  By performing a numerical study of
the late-time solutions of the proposed model, we found a unique set
of values ($\zeta, \xi$) such that its asymptotic limit describes the
correlations found in these numerical simulations.

Figure~\ref{fig:stresses_vs_emagnetic} shows the correlations between
the $r\phi$-com\-po\-nents of the Maxwell and Reynolds stresses and
the mean magnetic energy density found during the saturated state in
numerical simulations \cite{HGB95-96}. In our model, this ratio
depends only on the parameter $\zeta$ and the shear $q$ (held fixed at
$q=3/2$ in the simulations). It is evident that, in the numerical
simulations, the ratios of the stresses to the magnetic energy density
are also practically independent of any of the initial parameters in
the problem that determine the magnetic energy density during
saturation (i.e., the $x$-axis in the plot). This is indeed why we
required for our model of saturation to preserve the stress ratios
that are determined during the exponential phase of the MRI. Assigning
the same fractional uncertainty to all the numerical values for the
stresses, our model describes both correlations simultaneously for
$\zeta=0.3$.

Figure~\ref{fig:saturator_predictor} shows the mean magnetic energy
density in terms of a saturation predictor found in numerical
simulations \cite{HGB95-96}. For $\zeta=0.3$ and assigning the same
fractional uncertainty to all the numerical values for the stresses,
we obtain the best fit for $\xi=11.3$. These values for the parameters
complete the description of our model.

In summary, in this {\it Letter} we developed a local model for the
evolution and saturation of the Reynolds and Maxwell stresses in
MRI-driven turbulent flows. The model is formally complete when
describing the initial exponential growth and pumping of the
MRI-driven stresses and, thus, satisfies, by construction, all the
mathematical requirements described by Ogilvie~\cite{Ogilvie03}.
Although it is based on the absolute minimum physics (shear, uniform
$\bar{B}_z$, $2D$-fluctuations) for the MRI to be at work, the model
is able, in its asymptotic limit, to recover successfully the
correlations found in three-dimensional local numerical simulations
\cite{HGB95-96}.

Finally, the local model described in this {\em Letter} contains an
unexpected feature. The mean magnetic field in the vertical direction
couples the Reynolds and Maxwell stresses via the correlations between
the fluctuations in the velocity field and the fluctuating currents
generated by the perturbations in the magnetic field.  These second
order correlations, that we denoted by the tensor $\bar{W}_{ij}$, play
a crucial role in driving the exponential growth of the mean stresses
and energy densities observed in numerical simulations. To our
knowledge this is the first time that their relevance has been pointed
out in either the context of dynamo theory or MRI-driven turbulence.

\bigskip

We thank Eric Blackman, Gordon Ogilvie, Jim Stone, and an anonymous
referee for useful discussions. MEP acknowledges the hospitality of
the Institute for Advanced Study during part of this work.  MEP is
supported through a Jamieson Fellowship at the Astronomy Department at
the UA. This work was partially supported by NASA grant NAG-513374.


\bibliographystyle{apsrev}

\end{document}